\pdfoutput=1 
\documentclass[twocolumn,times]{aastex63}
\usepackage{graphicx}
\usepackage{mathtools,upgreek} 
\usepackage{natbib}
\usepackage{amsmath}
\usepackage{mathptmx}
\usepackage[version=4]{mhchem}
\hypersetup{pdfauthor={Lincowski et al.},pdftitle={Claimed detection of PH3 in the clouds of Venus is consistent with mesospheric SO2}}

\newcommand{\um}{$\upmu$m}

\newcolumntype{C}[1]{>{\centering\arraybackslash}m{#1}}

\received{October 21, 2020}
\revised{January 18, 2021}
\accepted{January 21, 2021}
\submitjournal{ApJL}

\newif\iftwocol
\twocolfalse

\begin{document}

\title{Claimed detection of \ce{PH3} in the clouds of Venus is consistent with mesospheric SO$_2$}

\author[0000-0003-0429-9487]{Andrew P. Lincowski}
\affiliation{Department of Astronomy and Astrobiology Program, University of Washington, Box 351580, Seattle, Washington 98195, USA}
\affiliation{NASA Nexus for Exoplanet System Science, Virtual Planetary Laboratory Team, Box 351580, University of Washington, Seattle, Washington 98195, USA}

\author[0000-0002-1386-1710]{Victoria S. Meadows}
\affiliation{Department of Astronomy and Astrobiology Program, University of Washington, Box 351580, Seattle, Washington 98195, USA}
\affiliation{NASA Nexus for Exoplanet System Science, Virtual Planetary Laboratory Team, Box 351580, University of Washington, Seattle, Washington 98195, USA}
\correspondingauthor{Victoria S. Meadows}
\email{meadows@uw.edu}

\author[0000-0002-4573-9998]{David Crisp}
\affiliation{NASA Nexus for Exoplanet System Science, Virtual Planetary Laboratory Team, Box 351580, University of Washington, Seattle, Washington 98195, USA}
\affiliation{Jet Propulsion Laboratory, California Institute of Technology, Earth and Space Sciences Division, Pasadena, California 91011, USA}

\author[0000-0001-8379-1909]{Alex B. Akins}
\affiliation{Jet Propulsion Laboratory, California Institute of Technology, Instruments Division, Pasadena, California 91011, USA}

\author[0000-0002-2949-2163]{Edward W. Schwieterman}
\affiliation{NASA Nexus for Exoplanet System Science, Virtual Planetary Laboratory Team, Box 351580, University of Washington, Seattle, Washington 98195, USA}
\affiliation{Department of Earth and Planetary Sciences, University of California, Riverside, CA 92521 USA}
\affiliation{Blue Marble Space Institute of Science, Seattle, WA, USA}

\author[0000-0001-6285-267X]{Giada N. Arney}
\affiliation{NASA Nexus for Exoplanet System Science, Virtual Planetary Laboratory Team, Box 351580, University of Washington, Seattle, Washington 98195, USA}
\affiliation{NASA/Goddard Space Flight Center, Greenbelt, MD 20771, USA}

\author[0000-0001-8212-3036]{Michael L. Wong}
\affiliation{Department of Astronomy and Astrobiology Program, University of Washington, Box 351580, Seattle, Washington 98195, USA}
\affiliation{NASA Nexus for Exoplanet System Science, Virtual Planetary Laboratory Team, Box 351580, University of Washington, Seattle, Washington 98195, USA}

\author[0000-0003-3962-8957]{Paul G. Steffes}
\affiliation{School of Electrical and Computer Engineering, Georgia Institute of Technology, Atlanta, GA 30332-0250, USA}

\author[0000-0003-1225-6727]{M. Niki Parenteau}
\affiliation{NASA Nexus for Exoplanet System Science, Virtual Planetary Laboratory Team, Box 351580, University of Washington, Seattle, Washington 98195, USA}
\affiliation{MS 239-4, Space Science Division, NASA Ames Research Center, Moffett Field, CA, USA}

\author[0000-0003-0354-9325]{Shawn Domagal-Goldman}
\affiliation{NASA Nexus for Exoplanet System Science, Virtual Planetary Laboratory Team, Box 351580, University of Washington, Seattle, Washington 98195, USA}
\affiliation{NASA/Goddard Space Flight Center, Greenbelt, MD 20771, USA}

\begin{abstract}

The observation of a 266.94 GHz feature in the Venus spectrum has been attributed to \ce{PH3} in the Venus clouds, suggesting unexpected geological, chemical or even biological processes. Since both \ce{PH3} and \ce{SO2} are spectrally active near 266.94 GHz, the contribution to this line from \ce{SO2} must be determined before it can be attributed, in whole or part, to \ce{PH3}. An undetected \ce{SO2} reference line, interpreted as an unexpectedly low \ce{SO2} abundance, suggested that the 266.94 GHz feature could be attributed primarily to \ce{PH3}. However, the low \ce{SO2} and the inference that \ce{PH3} was in the cloud deck posed an apparent contradiction. Here we use a radiative transfer model to analyze the \ce{PH3} discovery, and explore the detectability of different vertical distributions of \ce{PH3} and \ce{SO2}. We find that the 266.94 GHz line does not originate in the clouds, but above 80 km in the Venus mesosphere. This level of line formation is inconsistent with chemical modeling that assumes generation of \ce{PH3} in the Venus clouds. Given the extremely short chemical lifetime of \ce{PH3} in the Venus mesosphere, an implausibly high source flux would be needed to maintain the observed value of 20$\pm$10 ppb. We find that typical Venus \ce{SO2} vertical distributions and abundances fit the JCMT 266.94 GHz feature, and the resulting \ce{SO2} reference line at 267.54 GHz would have remained undetectable in the ALMA data due to line dilution. We conclude that nominal mesospheric \ce{SO2} is a more plausible explanation for the JCMT and ALMA data than \ce{PH3}.

\end{abstract}

\section{Introduction}

\citet{Greaves2020} recently attributed a 266.94~GHz (1.123~mm) line observed in the Venus spectrum to $\sim$20~ppb of phosphine (\ce{PH3}) absorbing above 56~km altitude, in the upper clouds. In the strongly-oxidizing Venus atmosphere,  \ce{PH3} formation is disfavored and its destruction is enhanced, leading \citet{Greaves2020} to argue that its presence in the clouds points to unknown geological, chemical or even biological processes. The discovery team identified no viable abiotic production mechanism for \ce{PH3} in the Venus atmosphere \citep{Greaves2020,Bains2020}, and so a biological origin was considered. \ce{PH3} has been proposed as a potential biosignature in terrestrial planet atmospheres \citep{Sousa-Silva2020} due to its association with decaying organic matter \citep{Glindemann2005}, and significant---presumed biological---fluxes from marine environments on Earth \citep{Zhu2007}. However, the specific mode of biological production of \ce{PH3} remains uncertain and is still vigorously debated \citep{Roels2001biological}, with no known direct metabolic pathway \citep{Roels2005}.  

The identification of \ce{PH3} in the Venus clouds was made using multiple observations of a single spectral feature at 266.94 GHz, where both \ce{PH3} (266.944~GHz) and \ce{SO2} (266.943~GHz) have absorption lines \citep{Greaves2020}. After the initial detection using coadded spectra from the James Clark Maxwell Telescope (JCMT), which were taken over 5 nights between 2017 June 9--16, follow-up observations were made with the Atacama Large Millimeter Array (ALMA) on 2019 March 5. The latter dataset included simultaneous narrow-band (0.1171875~GHz) and wide-band (1.875~GHz) observations, centered on the Venus rest-frame \ce{PH3} frequency. The 266.94~GHz line, seen in the JCMT data at a S/N of 4.3 (\citealt{Greaves2020}; although this detection significance has been subsequently called into question, \citealt{Thompson2020ph3}), was also detected in the ALMA narrowband and wideband datasets at higher significance than in the JCMT data \citep{Greaves2020}, although a subsequent reanalysis of the ALMA data also suggests a less significant detection, with a correspondingly lower inferred abundance of \ce{PH3} \citep{Greaves2020MA}.  Assuming a uniform mixing ratio for the \ce{PH3}, \citet{Greaves2020} derive an abundance of 20 ppb from the JCMT observations, and calculate an emission weighting function peaked at 56 km.  They therefore conclude that the \ce{PH3} absorption feature was sourced primarily from within the Venus clouds.  However, as \citet{Greaves2020} point out, with a FWHM of 4--5 km~s$^{-1}$, this line could potentially contain contributions from both \ce{PH3} and \ce{SO2}, as the \ce{SO2} line center is only +1.3~km~s$^{-1}$ from the \ce{PH3} line center. 

Consequently, the \ce{PH3} line identification is strongly dependent on accurately estimating and excluding a potentially-significant contribution from \ce{SO2}, which, after the bulk atmospheric gases \ce{CO2} and \ce{N2}, is the third most abundant gas in the Venus atmosphere. \citet{Greaves2020} attempted to quantify the \ce{SO2} contribution to the observed 266.94~GHz feature by searching the ALMA wide-band observations for the nearby, stronger \ce{SO2} $J_{K_a,K_c} = 13_{3,11}\leftarrow13_{2,12}$ 267.537~GHz line, but did not detect it (see their Figure 4a). Instead, they estimated a 10~ppb upper limit for \ce{SO2}, based on potentially large spectral ``ripples", artifacts in the data induced by interferometric response to Venus as a bright, extended source. \citet{Greaves2020} also noted that the $\leq10$~ppb value was comparable to a 346.652~GHz ALMA Venus \ce{SO2} measurement of $16.5\pm4.6$~ppb, which was taken in 2011 \citep{Piccialli2017}.  However, the \citet{Piccialli2017} observation was sensitive to \ce{SO2} at 85~km altitude \citep{Piccialli2017} in the Venus mesosphere (which extends from 65--120~km), and not to the middle/upper cloud deck (53--61~km).  The $\leq10$~ppb constraint derived from the non-detection implied a maximum 10\% contribution from \ce{SO2} to the 266.94~GHz absorption band depth, and a shift in the observed line centroid of no more than 0.1~km~s$^{-1}$. \citet{Greaves2020} concluded that \ce{SO2} had been ruled out as a significant contaminant for the putative \ce{PH3} line.  Conversely, they argued that the 266.94 GHz line could not be explained solely by \ce{SO2}, because the corresponding reference lines would be significantly stronger than the $-0.0006$~l:c (line-to-continuum) ratio limit set by the spectral ripples, and yet the reference lines were not detected. 

\begin{figure*}[tbh]
    \centering
    \includegraphics[width=0.33\textwidth]{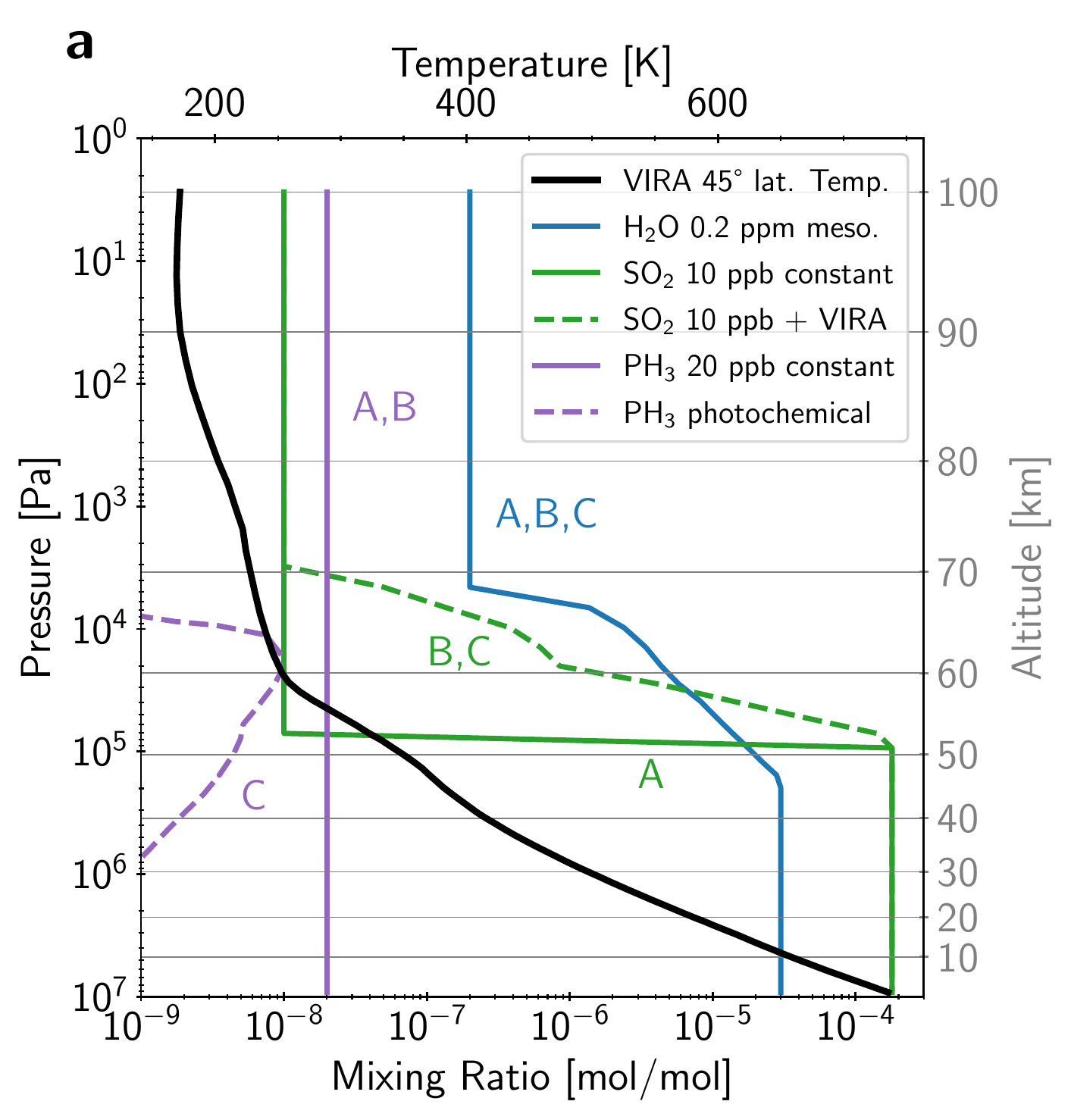}
    \includegraphics[width=0.33\textwidth]{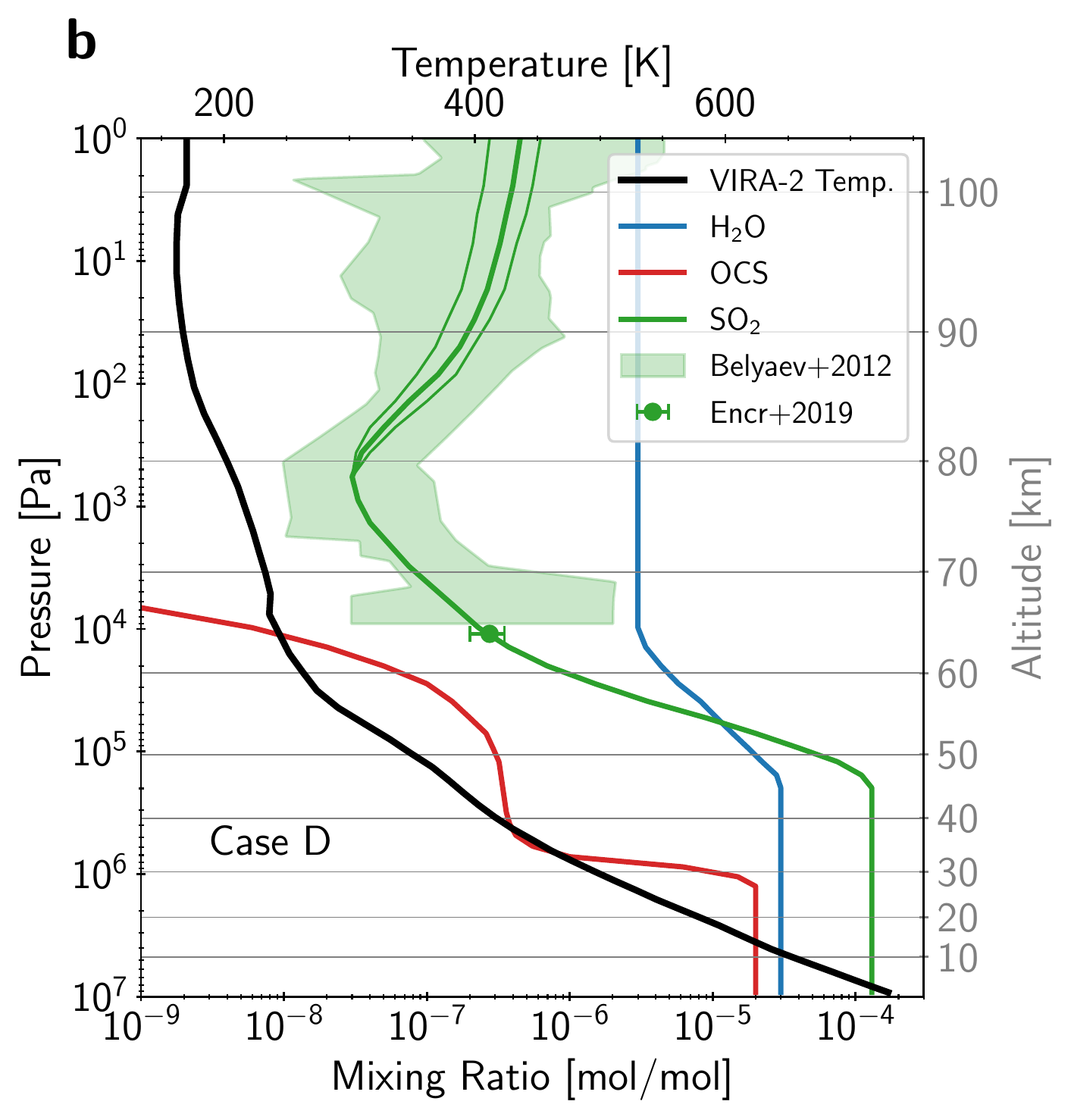}
    \includegraphics[width=0.33\textwidth]{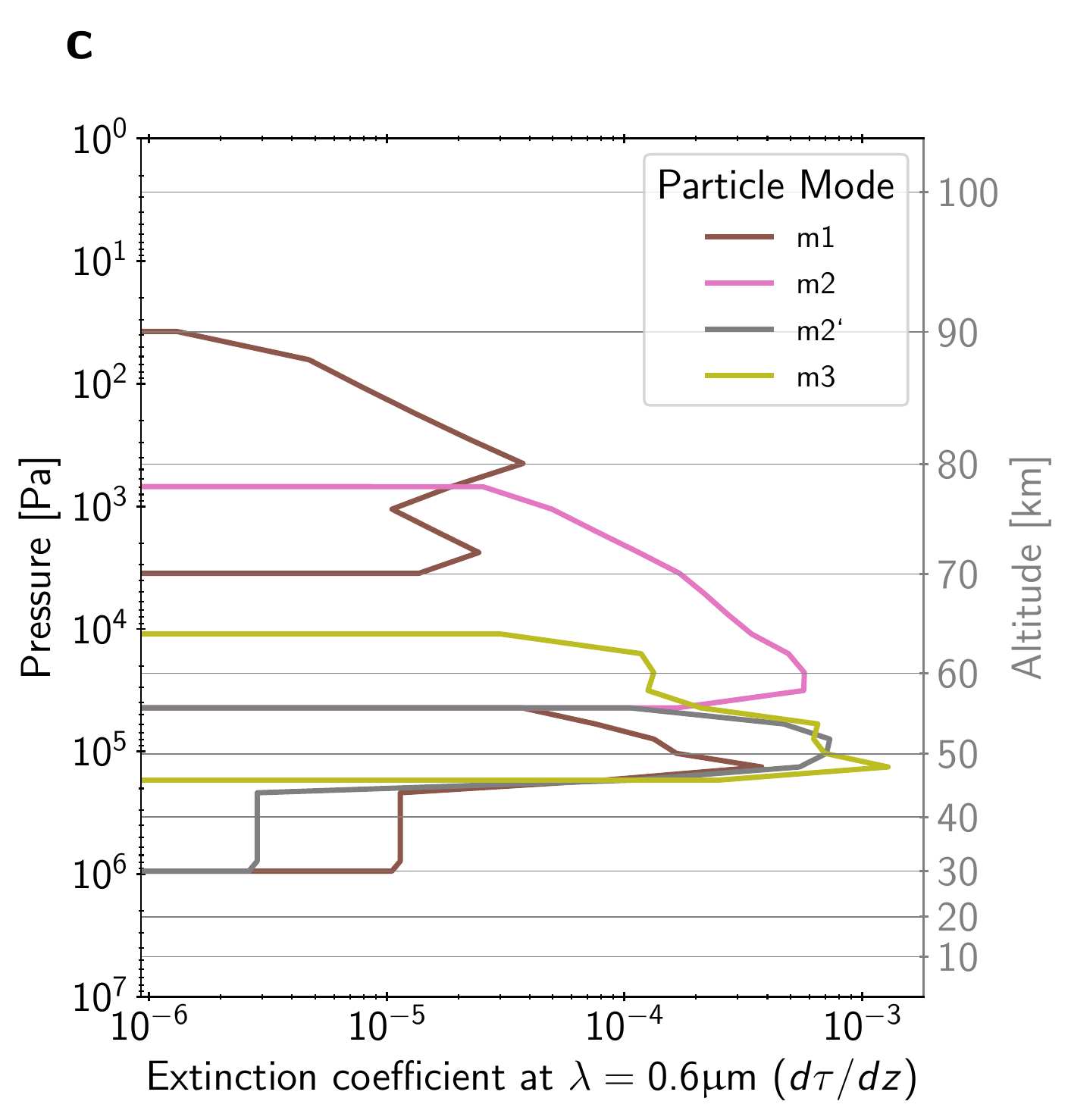}
    \caption{Atmospheric structures for Venus used in our spectral modeling cases. \textit{Panel~a}: Temperature and vertical profiles for Cases A--C, which use parameters assumed/derived by \citet{Greaves2020}. Temperature profile (black line) is for VIRA 45$^\circ$ latitude (\citealt{Seiff1985}), and vertical profiles are shown for \ce{PH3} (solid, Cases A, B; dashed Case C), \ce{SO2} (green), and \ce{H2O} (blue).  \textit{Panel~b}: Temperature and vertical profiles for our Case D best fit to the JCMT 266.94 GHz line. The temperature profile (black line) is from VIRA-2 \citep{Moroz1997VIRA2}. The nominal gas mixing ratios for \ce{H2O} (blue line) is based on VIRA values \citep{vonZahn1985composition} updated for the lower atmosphere \citep{debergh2006} but have also been modified slightly as described in \S~\ref{sec:methods}. For OCS (red line), the profile is constructed based on recent measurements by \citet{Krasnopolsky2010so2} and \citet{Arney2014}, and by the surface abundance in the lower atmosphere model by \citet{Krasnopolsky2013}. For \ce{SO2} (green line), we fit the 266.94~GHz line guided by the vertical profile of \citet{belyaev2012vertical} in the mesosphere and upper cloud, and consistent with a suite of SOIR and SPICAV UV \ce{SO2} measurements taken from 2007--2008 (green shaded region). This profile passes through the cloud-top \ce{SO2} measurement (200--350~ppb) obtained by \citet{Encrenaz2019hdoso2} in July 2017 (green data point), one month after the \citet{Greaves2020} JCMT observations. We use 130~ppm in the lower atmosphere \citep{marcq2008latitudinal} and generated a profile between the lower atmosphere and cloud tops. \textit{Panel~c}: optical depth extinction profiles (optical depth per meter at a wavelength of 0.6~\um{}) for the Venus cloud particle modes: m1 (haze), m2, m2', and m3 \citep{Crisp1986}. The clouds are defined via optical depth considerations to span approximately 48--70~km ($3\times10^3$--$1.3\times10^5$~Pa).}
    \label{fig:mixes}
\end{figure*}

Because the non-detection of \ce{SO2} by \citet{Greaves2020} supports a corresponding low inferred abundance, and a low contamination fraction for the 266.94 GHz line, it is the key piece of evidence supporting the \ce{PH3} line identification at 266.94~GHz--- and so it warrants closer scrutiny. There is an apparent contradiction between the inferred altitudes that the \ce{PH3} feature probed, and the \ce{SO2} abundance constraint. If the putative \ce{PH3} (266.94~GHz) absorption is sensitive to altitudes near 56~km, and thus probes the Venus middle and upper cloud, then the 267.94~GHz \ce{SO2} reference line should also originate from this altitude range, since it has similar line strength and amount of underlying continuum absorption. Data and modeling estimates place the \ce{SO2} abundance near 1--5 ppm at 60~km in the upper cloud, which should increase with depth to match the higher $\sim$130~ppm measured below the cloud deck \citep{Zasova1993,Krasnopolsky2012,Zhang2012,belyaev2012vertical,marcq2008latitudinal,Arney2014,Encrenaz2019hdoso2}. Previous measurements therefore suggest that the inferred disk-averaged $<10$ ppb of \ce{SO2} is anomalously low, especially if the observations probe within the clouds.  Assuming similar spatial distribution of the two gases, for an inferred \ce{SO2} abundance at 56~km of 10ppm, and the 10 ppb \ce{PH3} abundance of \citet{Greaves2020}, the \ce{SO2} contribution to the observed line would exceed that from \ce{PH3} by two orders of magnitude \citep{krasnopolsky2020}.

If the observations were instead sensitive to the mesospheric levels above the clouds, as is the case for higher frequency ALMA observations \citep{sandor2010sulfur,Encrenaz2015venus}, then the inferred Venus \ce{SO2} abundance would be closer to, but still lower than previously measured levels \citep{sandor2010sulfur,Encrenaz2015venus,Piccialli2017,vandaele2017sulfur}. While the abundance of \ce{SO2} above the clouds is known to vary significantly over time \citep{esposito1988sulfur,encrenaz2012hdo,Encrenaz2019hdoso2} with a minimum observed around 10-100~ppb at $\sim$80~km, the abundances in the mesosphere have been measured to be in the range 10~ppb to 10~ppm  \citep{Krasnopolsky2010so2,belyaev2012vertical,vandaele2017sulfur}. A planet-wide decrease from a higher cloud-top \ce{SO2} abundance in 2006 to a low in 2014 of 30 ppb was also observed, but more recent observations from 2016 through September 2018, which span the \citet{Greaves2020} JCMT observation, show a strong increase to typical cloud-top values of several hundred ppb of \ce{SO2} \citep{Encrenaz2019hdoso2}.

While line absorption occurring predominantly within the mesosphere would make the non-detection and inferred low abundance of \ce{SO2} more plausible, it would also suggest that the line attributed to \ce{PH3} was formed at mesospheric levels. Consequently, the 266.94~GHz line would not be sensitive to, and so not able to confirm, the presence of \ce{PH3} in the Venus clouds---potentially weakening support for a biological origin.  The presence of 20~ppb of mesospheric \ce{PH3} would require an extremely large source flux due to photolysis and reactions with radical species, including Cl and H, that result in a sub-second lifetime for \ce{PH3} in the Venus mesosphere  \citep[][their Fig. 2]{Bains2020}. Indeed, the vertical distribution predicted using photochemical-kinetics studies with a cloud source of \ce{PH3} indicates a sharply reduced mesospheric abundance of \ce{PH3} ($<0.001$~ppb) alongside significant ($>100$~ppb below 95~km) \ce{SO2} (\citealt{Greaves2020}, extended data figure 9; \citealt{Bains2020}).

To explore the potential contradictions posed by the \citet{Greaves2020} \ce{PH3} observations, and to verify the source region for the 266.94 GHz absorption, here we use a radiative transfer model of the Venus atmosphere to simulate the impact on the Venus millimeter-wavelength spectrum of different abundances and vertical distributions of \ce{PH3} and \ce{SO2}, including those proposed by \citet{Greaves2020} and \citet{Bains2020}.

\section{Methods} \label{sec:methods}

To generate synthetic millimeter-wavelength spectra of Venus, we use SMART (Spectral Mapping Atmospheric Radiative Transfer), a 1D line-by-line, multi-stream, fully multiple-scattering radiative transfer model \citep{Meadows1996,Crisp1997}. SMART has been validated against observations of Solar System planets, with heritage modeling the Venus atmosphere \citep{Meadows1996,Arney2014,Robinson2018}. 

Our spectral simulations consist of Cases A--C, for which we generate spectra based on the mixing ratios and vertical profiles used and derived by \citet{Greaves2020}, and our best fit model, Case D, which does not contain \ce{PH3} and uses constraints from additional Venus observations (Figure \ref{fig:mixes}). Cases A--C include \ce{CO2}, \ce{SO2}, \ce{H2O}, and \ce{PH3} and use the VIRA 45$^\circ$ latitude temperature profile \citep{Seiff1985}. To match the \ce{H2O} estimate of \citet{Greaves2020}, we use the \citet{debergh2006} \ce{H2O} profile but reduced to 0.2~ppm above 68~km. For \ce{SO2}, we use the \citet{debergh2006} compilation below 100~km for cases B and C, but reduced to 10~ppb above 70~km, and for case A we maintain 10~ppb down through the cloud deck to 53~km. For \ce{PH3}, we use a uniformly mixed 20~ppb profile for cases A and B, and the photochemical profile from \citet{Greaves2020} (their figure ED7) for case C.

For our best-fit scenario, Case D, we do not include \ce{PH3} and use the \citet{debergh2006} update to the VIRA below 100~km and more recent observations where available. We use the VIRA-2 temperature profile \citep{Moroz1997VIRA2}. For \ce{H2O}, we use 30~ppm below the cloud deck \citep[][ and references therein]{debergh2006}, and we assume 3~ppm above the cloud deck \citep{Krasnopolsky2013dh,Cottini2015,Piccialli2017}. For \ce{SO2}, we use 130~ppm below the cloud deck \citep{Gelman1979,Bezard1993,debergh2006,marcq2008latitudinal,Arney2014}, decreasing with increasing altitude to the July 2017 observation of $\sim$275~ppb at 64~km \citep{Encrenaz2019hdoso2}, which was measured within a month of the \citet{Greaves2020} JCMT data. In the mesosphere, we fit the \ce{SO2} profile to the observed feature at 266.94~GHz guided by the vertical profile fit to  2007--2008 data from \citet{belyaev2012vertical}, which is consistent with the cloud-top \ce{SO2} abundance observed in July, 2017 \citep[see][]{Encrenaz2019hdoso2}. Long-term monitoring has shown that 2007--2008 and 2017--2018 were similar maximum periods of global mesospheric \ce{SO2} abundance \citep{Encrenaz2019hdoso2}, although short-term temporal variability within these secular changes can be orders of magnitude \citep{Belyaev2017}. We prescribe the OCS profile guided by recent measurements \citep{Krasnopolsky2010so2,Arney2014} and models \citep{Zhang2012,Krasnopolsky2012,Krasnopolsky2013,Lincowski2018T-1}.
We adopt the same aerosol properties, modes, and optical depth profiles as \citet{Arney2014}, which originate from \citet{Crisp1986}. Temperature and gas profiles, and aerosol optical depths, are shown in Figure~\ref{fig:mixes}.

Absorption cross-sections associated with vibrational-rotational transitions are calculated using a line-by-line model, LBLABC \citep[see][for details]{Meadows1996,Crisp1997}, with the HITRAN2016 line database \citep{Gordon2017} for all gases except \ce{CO2}, which is calculated from the extensive Ames line database \citep{Huang2017}. Because these line lists assume terrestrial isotopic abundance, we use the methods described in \citet{Lincowski2019} to adjust the line list isotopologue abundances for \ce{H2O} to 200 times the D/H abundance compared to Earth, the standard value used in the literature for the Venus mesosphere \citep{Encrenaz2015venus}.
Collision-induced absorption data is used for \ce{CO2}-\ce{CO2} \citep{Gruszka1997}.

Data on the foreign broadening of gases by \ce{CO2} is not well-characterized, compared to broadening by air, but is more appropriate for Venus simulations.  To reproduce the results of \citet{Greaves2020}, we use their foreign broadening parameter for \ce{PH3} of 0.186~cm$^{-1}$~atm$^{-1}$, which they used to estimate \ce{PH3} as 20$\pm10$~ppb in the JCMT data. Because their broadening treatment for gases other than \ce{PH3} is not specified, we use the default HITRAN air broadening for cases A--C.
To fit the 266.94~GHz detection feature with \ce{SO2} in case D, we employ data for broadening by \ce{CO2}, as available. For \ce{SO2} and OCS, we use data for broadening by \ce{CO2} available in HITRAN \citep{Wilzewski2016,Gordon2017}. Although the \ce{SO2} broadening data are derived from a single line experiment \citep{Chandra1963so2_co2}, the parameters in the frequencies of interest are consistent with recent laboratory results by \citet{Bellotti2015}. The broadening values for our \ce{SO2} lines of interest are approximately 1.8--2.0$\times$ air broadening (i.e. $\gamma_{\ce{CO2}}\simeq0.17-0.19$~cm$^{-1}$~atm$^{-1}$). For HDO, we multiply the HITRAN air foreign broadening parameters by 2.4, which is consistent with this frequency range \citep{Sagawa2009}.

To better visualize individual line signal and compare to the published data, we processed our flux spectra to normalize the continuum. Because we are processing noiseless model results, we mask spectral intervals for individual lines and linearly interpolate the continuum across the interval. The line:continuum (l:c) spectra were determined by dividing the original model spectrum by the continuum and subtracting one.

\begin{figure}[tbh]
    \centering
    \includegraphics[width=0.66\columnwidth]{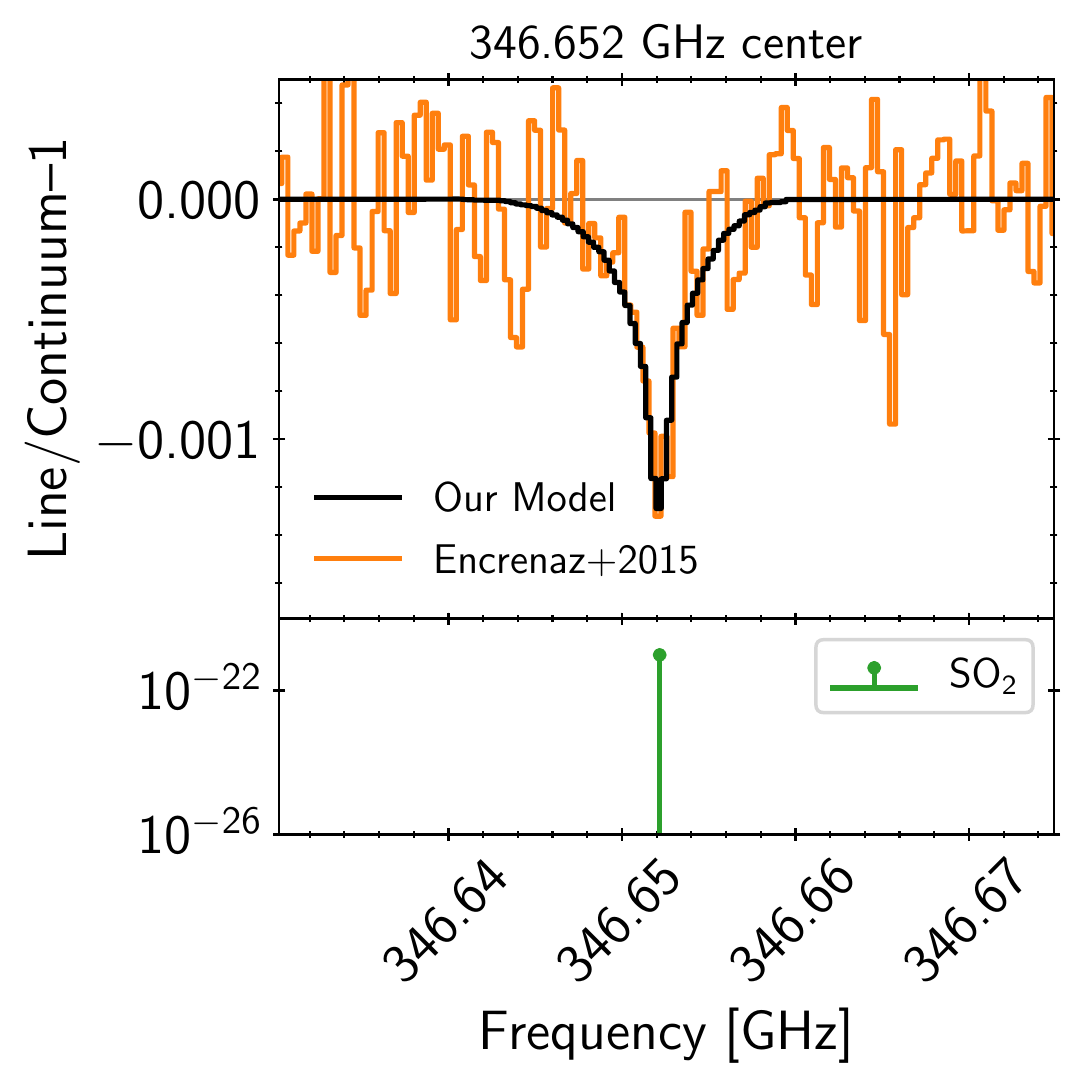}
    \caption{We demonstrate the validity of our model by fitting the 346.652~GHz \ce{SO2} line observed by \citet[][Fig.~19]{Encrenaz2015venus} in 2011 using their best-fit profile of no \ce{SO2} from 70--85~km and 10~ppb above 85~km, with all other modeling parameters specified as for our Case D (but also including CO from the \citet{debergh2006} compilation). \ce{SO2} absorption line strength in the bottom panel is given in units of cm$^{-1}$/(molecule cm$^{-2}$). This comparison shows our model and associated parameters are consistent with previous sub-mm observations of \ce{SO2}.    \label{fig:encr} }
\end{figure}

As an additional validation of our radiative transfer model and fit to the \citet{Greaves2020} JCMT data, we applied our model to simulate the line shape and peak intensity of the 346.65~GHz late-2011 observation of \citet{Encrenaz2015venus}, using their \ce{SO2} profile of 10~ppb from 86--100 km, and obtained an excellent fit to the data (see Fig.~\ref{fig:encr}).

\section{Results}

\begin{figure*}
    \centering
    \includegraphics[width=0.9\textwidth]{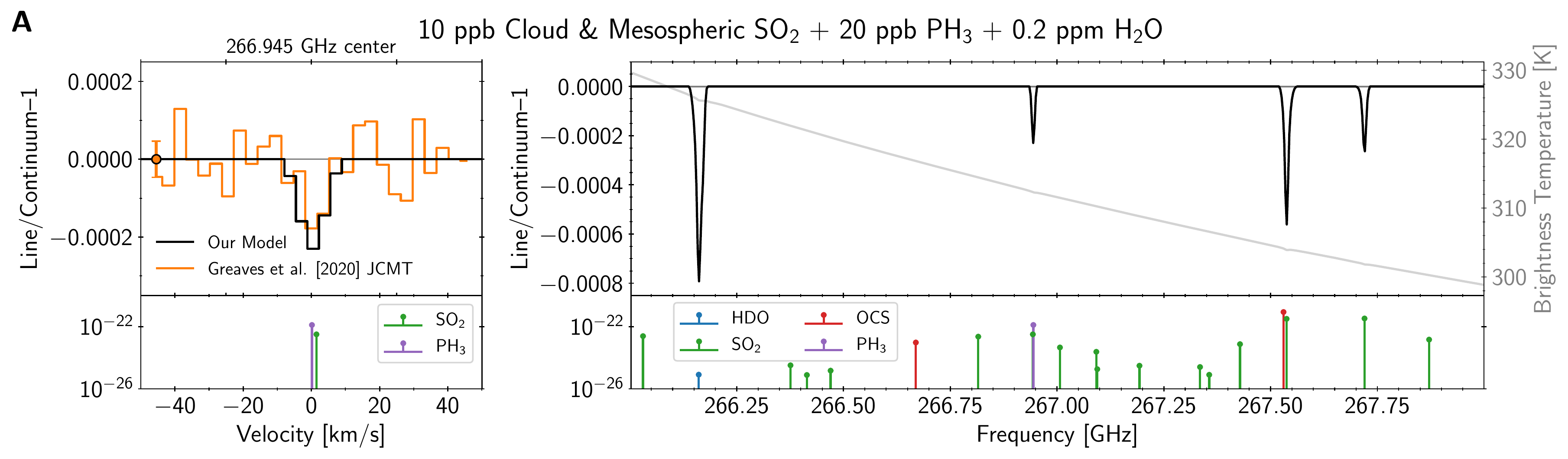}
    \includegraphics[width=0.9\textwidth]{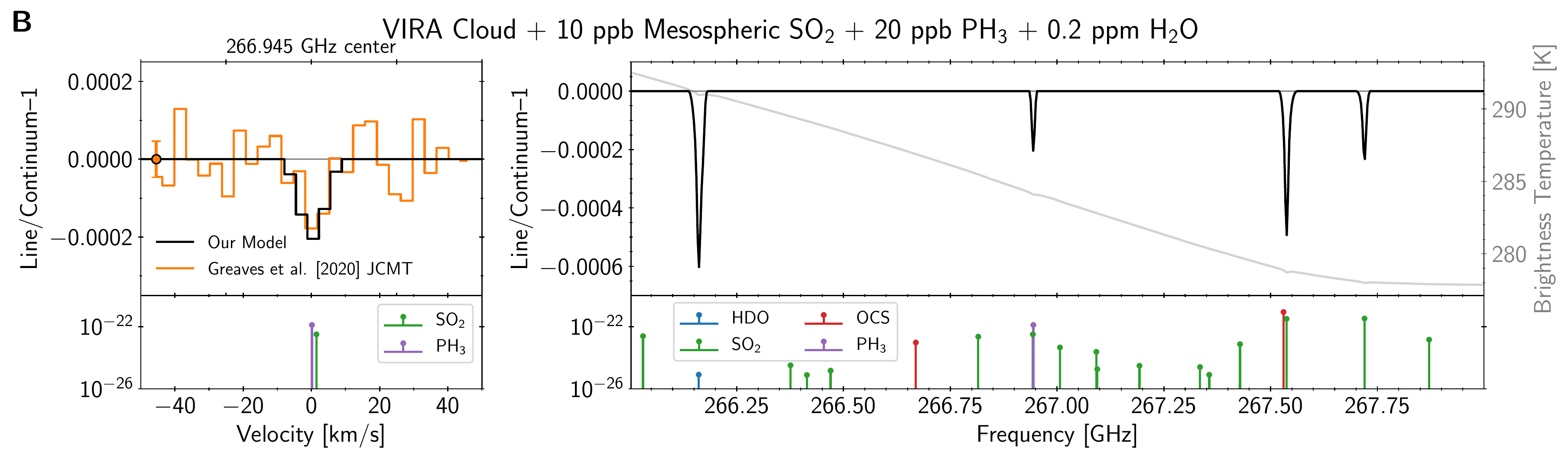}
    \includegraphics[width=0.9\textwidth]{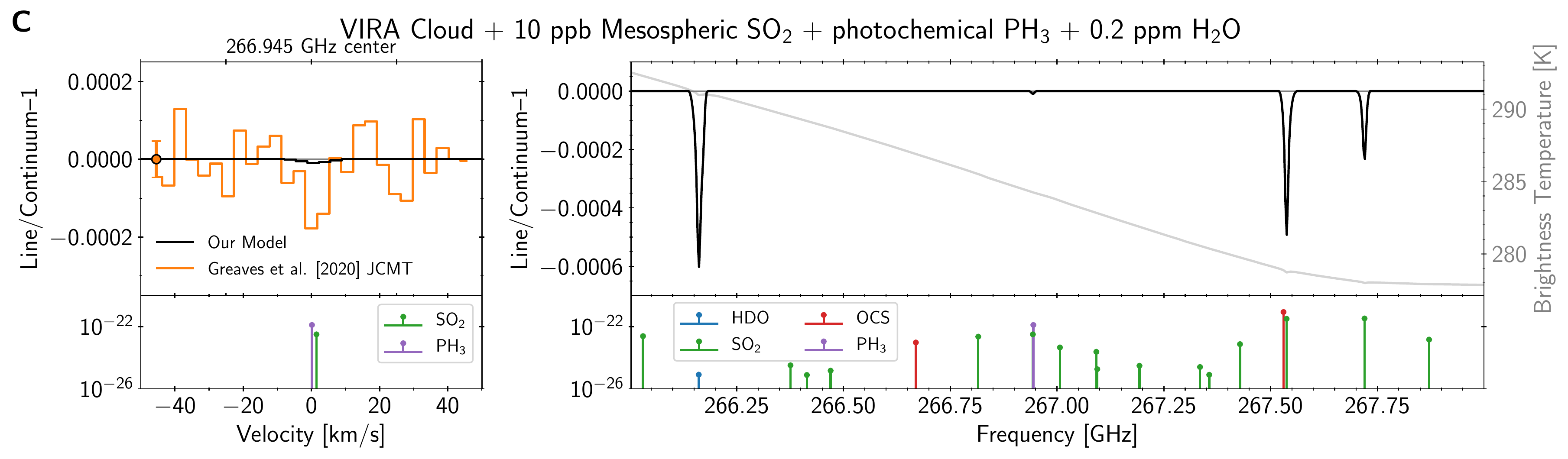}
    \includegraphics[width=0.9\textwidth]{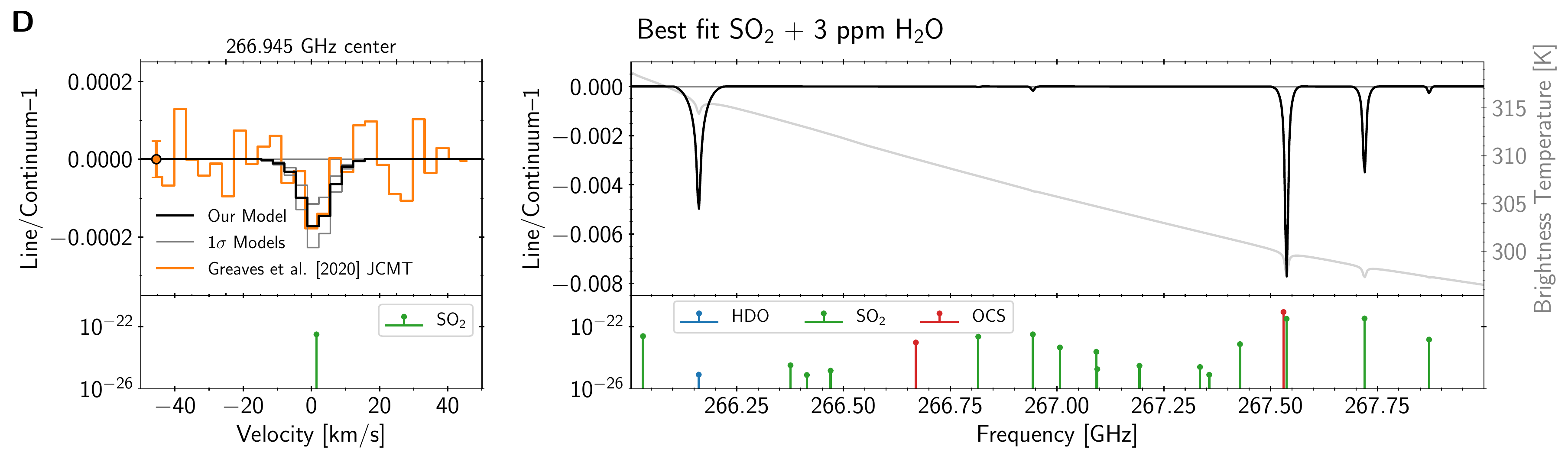}
    
    \caption{\footnotesize Venus spectral simulations for different \ce{PH3} and \ce{SO2} abundances and vertical profiles, including brightness temperature spectra (grey lines) to show the continuum source, and absorption line strengths (lower panels in each case) in units of cm$^{-1}$/(molecule cm$^{-2}$). For each case, the left hand panel shows the corresponding fit to the 266.94~GHz line, the right panel shows the 266--268~GHz spectrum, including the \ce{SO2} reference line at 267.54~GHz. \textit{Case A}: modified VIRA temperature and gas profiles with uniformly mixed 20~ppb \ce{PH3} and 10 ppb \ce{SO2} down through the cloud deck (c.f. Figure \ref{fig:mixes} Panel a).  \textit{Case B}: Case A but with the VIRA \ce{SO2} profile in the cloud deck up to 70~km instead of evenly mixed at 10 ppb. \textit{Case C}: VIRA and \ce{SO2} profile as in Case B, but using the photochemically self-consistent profile for \ce{PH3} from \citet{Greaves2020} (ED Fig. 9). \textit{Case D}: VIRA-2 temperature profile, no \ce{PH3}, and using a vertically-resolved  \ce{SO2} profile derived from a suite of spacecraft and ground-based measurements, with a mesospheric profile that increases from 30~ppb at 78~km to 400$\pm$150~ppb at 100~km (see Figure 1, panel b). Cases A and B demonstrate similar fits for \ce{PH3} to the the 266.94~GHz line as in \citet{Greaves2020}, and show lack of sensitivity to the vertical distribution of \ce{SO2} in the clouds. Case C demonstrates that the \ce{PH3} profile generated assuming a source in the Venus clouds is inconsistent with the observed 266.94~GHz line. Case D shows that we can fit the detection feature with no \ce{PH3} but with a typical Venus \ce{SO2} abundance, although this produces \ce{SO2} reference line features that are over 10 times stronger than the other cases.} 
    \label{fig:spectra}
\end{figure*}

To explore the spectral impacts of different abundances and vertical profiles for \ce{PH3} and \ce{SO2}, we simulated spectra of Venus from 266 to 268~GHz.  This spectral range includes the \ce{HDO}, \ce{PH3} and \ce{SO2} line positions discussed in \citet{Greaves2020}, as well as OCS, which includes a transition at 267.530~GHz. 
We simulated spectra for cases with the abundances determined by \citet{Greaves2020} and vertical profiles determined by previous measurements of the Venus atmosphere (Figure \ref{fig:mixes}). Line-to-continuum (l:c) spectra generated at 0.0001~cm$^{-1}$ (3~MHz) resolution are shown in Figure~\ref{fig:spectra}, along with the emission brightness temperature in grey. The brightness temperatures demonstrate the effective altitude of continuum emission, and are directly correlated with \ce{SO2} abundance in the cloud deck between 54--57~km, depending on the case. Lower cloud \ce{SO2} abundance (10~ppb evenly-mixed) yields higher continuum emission from deeper in the atmosphere.

\subsection{Simulated Spectra}

For our Case A spectral simulation (Figure~\ref{fig:spectra}A) we assumed updated VIRA-derived profile (our Figure~\ref{fig:mixes}, see \citealt{vonZahn1985composition,debergh2006}) for all constituents except \ce{SO2} and \ce{PH3}. Following \citet{Greaves2020}, we assumed an evenly mixed abundance of 20~ppb \ce{PH3} and 10~ppb \ce{SO2} above 52~km altitude (near the base of the Venus cloud deck; green and purple dotted lines in Figure~\ref{fig:mixes}a). We also assumed their foreign broadening parameter for \ce{PH3} of 0.186~cm$^{-1}$~atm$^{-1}$. Our model produces a comparable fit to \citet{Greaves2020} for the 266.94~GHz line (c.f. their Figure~1). Additionally, with the evenly-mixed 10~ppb of \ce{SO2}, we also confirm that the 267.54~GHz \ce{SO2} line is below the spectral-ripple-inferred maximum limit on the l:c ratio ($-0.0006$). 

In our Case B simulation (Figure~\ref{fig:spectra}B), instead of assuming the low 10~ppb \ce{SO2} down through the cloud deck, we used the VIRA-derived profile such that the \ce{SO2} abundance increased with cloud depth (green dashed line in Figure~\ref{fig:mixes}b). At the 56~km level, the \ce{SO2} abundance is now closer to 20~ppm.  The increased \ce{SO2} opacity raises the emission layer to cooler levels of the atmosphere, as shown in the the brightness temperature difference between Cases A and B. This produces a small change in the \ce{SO2} continuum, which results in only marginal differences in the intensities of the 266.94~GHz \ce{PH3} line and the 267.54~GHz \ce{SO2} line, and the latter is still consistent with the maximum limit in sensitivity due to spectral ripple.  Thus the observed line intensities are largely insensitive to \ce{SO2} abundance within the clouds.

In our Case C simulation (Figure~\ref{fig:spectra}C), we again used 10~ppb \ce{SO2} in the mesosphere, increasing through the cloud deck (green dashed line in Figure \ref{fig:mixes}a). However, instead of \ce{PH3} evenly mixed throughout the atmosphere (as in Cases A and B), we used the photochemical profile for \ce{PH3} used to interpret the 266.94~GHz detection, as provided in \citet[][their ED~Figure~9]{Greaves2020}, and \citet[][reproduced as the purple dashed line in our Figure~\ref{fig:mixes}a]{Greaves2020}.  This distribution is derived from the assumption that \ce{PH3} production is concentrated within the cloud deck with abundance dropping rapidly in the upper cloud deck and mesosphere, and more slowly towards the surface.  The small absorption line present here at 266.94~GHz is due to \ce{SO2}---no \ce{PH3} absorption is visible in this spectral simulation. This indicates that the line core observation is not sensitive to \ce{PH3} in the cloud, and demonstrates that the assumed profile in the \citet{Greaves2020} and \citet{Bains2020} photochemical simulations are inconsistent with the JCMT observations.  

In our Case D simulation (Figure \ref{fig:spectra}D), we removed \ce{PH3} from our atmosphere and fit the JCMT detection feature at 266.94~GHz using \ce{SO2} alone. As described in \S\ref{sec:methods}, we used parameters for HDO, \ce{SO2}, and OCS foreign broadening by \ce{CO2}. We guided the mesospheric data fit for \ce{SO2} using Venus Express UV/IR occultation data from \citet{belyaev2012vertical}. This profile is consistent with cloud-top \ce{SO2} abundances measured by \citet{Encrenaz2019hdoso2} within a month of the \citet{Greaves2020} JCMT observations. Our best-fit \ce{SO2} profiles, fitting the observed line (black) and $\pm$1-$\sigma$ about the line (grey) are shown in Figure~\ref{fig:mixes}b (green curves), with \ce{SO2} increasing from 30~ppb at 78~km to 400$\pm$150~ppb at 100~km. These abundance profiles are well within the range of measurements compiled in \citet{belyaev2012vertical} and \citet{vandaele2017sulfur}. This simulation provides an excellent fit to the JCMT detection line without \ce{PH3}, and  predicts a pair of \ce{SO2} reference lines that have l:c ratios a factor of $\sim$10 higher than those seen in the previous simulations.

\subsection{Spectral Line Sensitivity}

\begin{figure*}[tbh]
    \centering
    \includegraphics[width=0.45\textwidth]{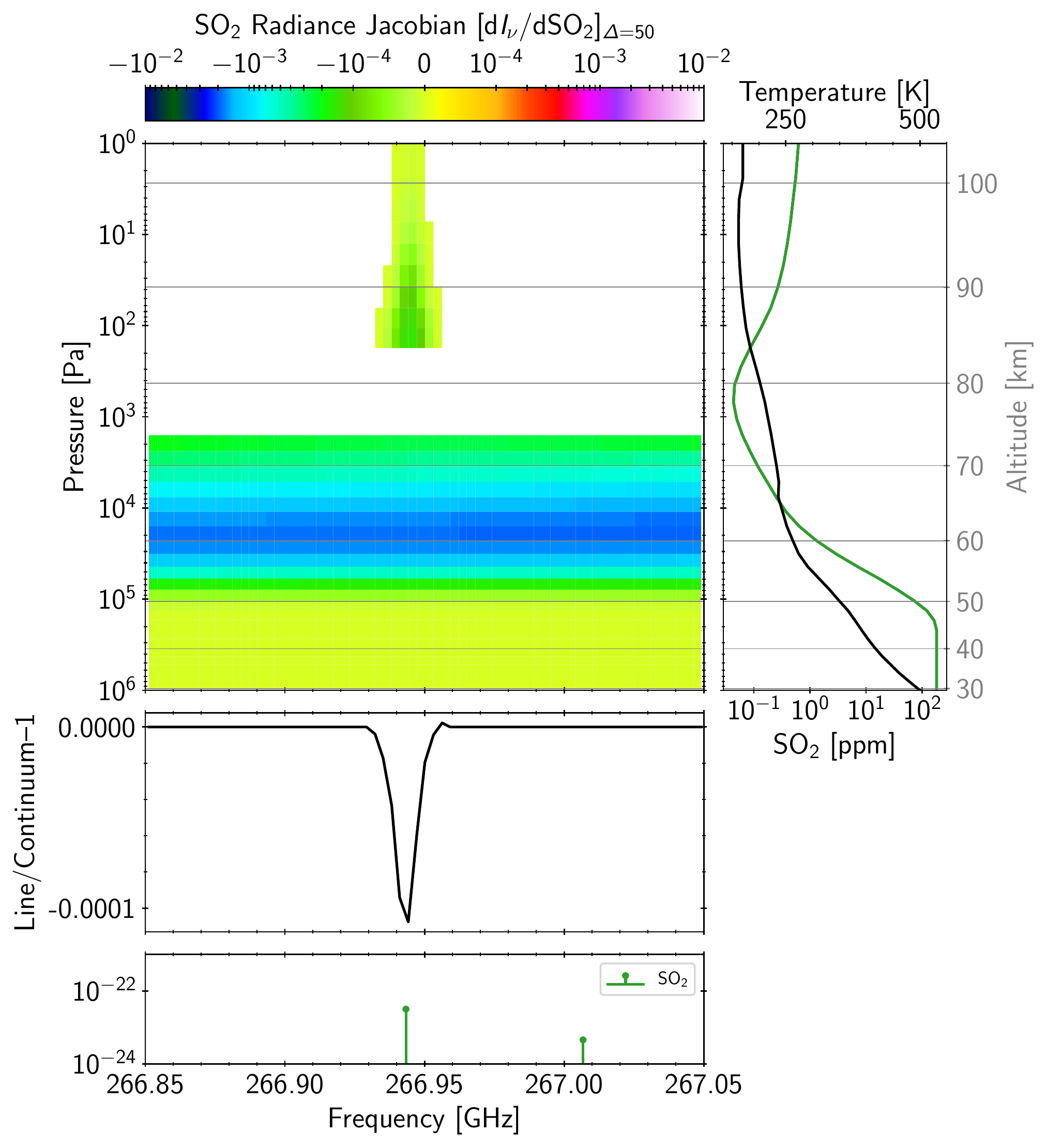}
    \includegraphics[width=0.45\textwidth]{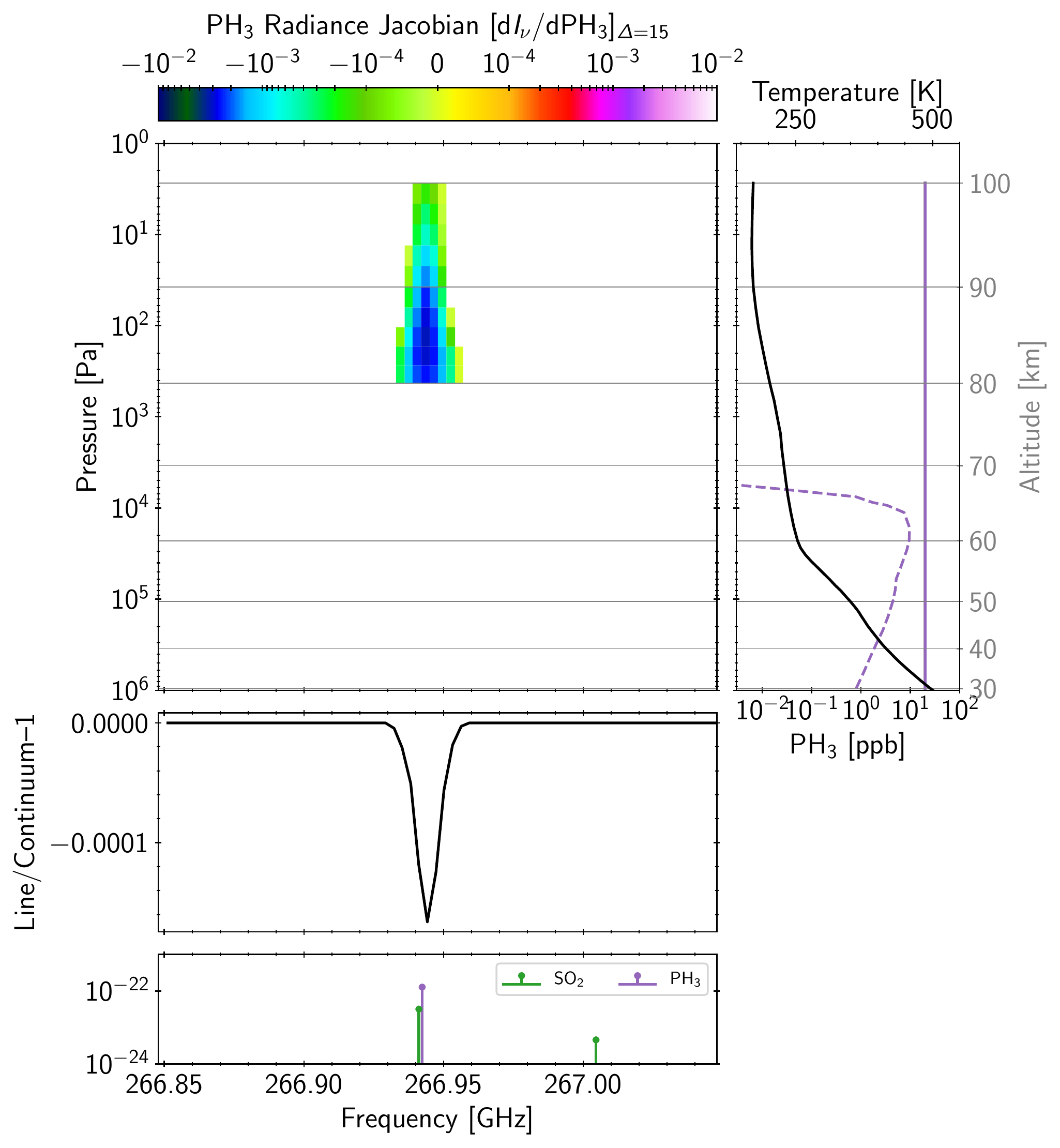}
    \caption{Radiance mixing ratio Jacobians ($dI_\nu/dr_\text{mix}$) as a function of layer pressure for the radiance streams at 21~degrees zenith angle, for the 266.94~GHz feature for \ce{SO2} (left) or \ce{PH3} (right). Absorption line strengths (lower panels in each case) are given in units of cm$^{-1}$/(molecule~cm$^{-2}$). The continuum originates from \ce{SO2} at $\sim6\times10^4$~Pa ($\sim$54~km, within the cloud deck), while the line cores for either species do not originate in the clouds (48-70~km) but at over 400~Pa (over 80~km) in the mesosphere. In the right panels in both plots, the temperature structure is given as a black line, while the colored lines denote \ce{SO2} (green) or \ce{PH3} (purple) mixing ratios. On the right, the evenly-mixed 20~ppb \ce{PH3} profile is shown with a solid line and the photochemical \ce{PH3} profile is shown with a dashed line.
    }
    \label{fig:jacobians}
\end{figure*}

To confirm the altitudinal sensitivity of the 266.94~GHz line for key \ce{PH3} and \ce{SO2} vertical profiles, we calculated radiance Jacobians, i.e. the increase in top-of-atmosphere radiance as a function of perturbations to the abundances for \ce{SO2} and \ce{PH3} at each layer of our model atmosphere (Figure~\ref{fig:jacobians}). The outgoing radiance will be most sensitive to regions of the atmosphere that contribute most to the spectral feature.  The Jacobians show that the observed line cores for both gases originate from atmospheric pressures only as deep as $\sim$400~Pa, corresponding to altitudes of $\ge$80~km, in the mesosphere. This absorption feature cannot be generated at levels within the cloud deck, where the background continuum emission originates. It must be generated well above this layer, where the absorbing gas is cooler and therefore absorbs more efficiently than it emits. The narrow width of the absorption line also suggests that it was formed at pressures substantially less than those of the cloud top (70~km, $\sim$3000~Pa).

\subsection{ALMA Line Dilution}

\begin{figure*}[tbh]
    \centering
    \includegraphics[scale=0.45]{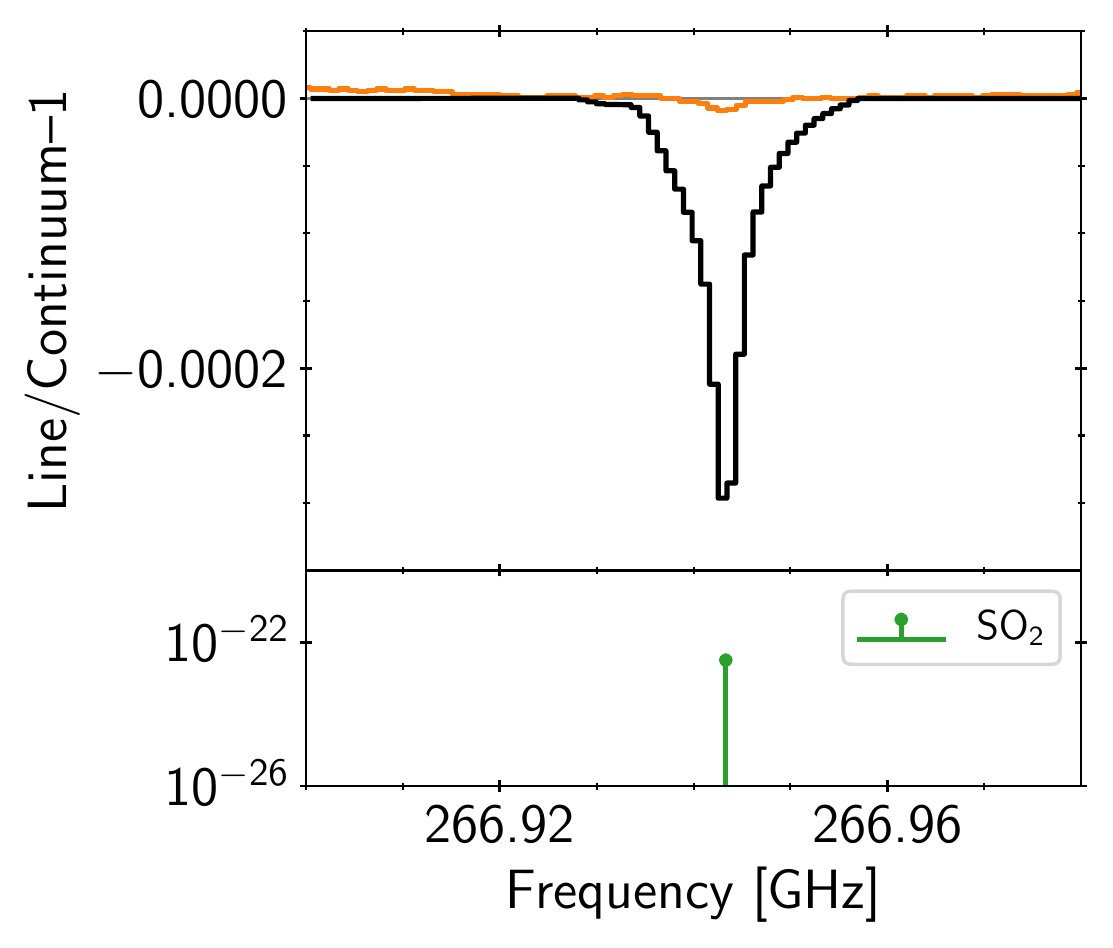}
    \includegraphics[scale=0.45]{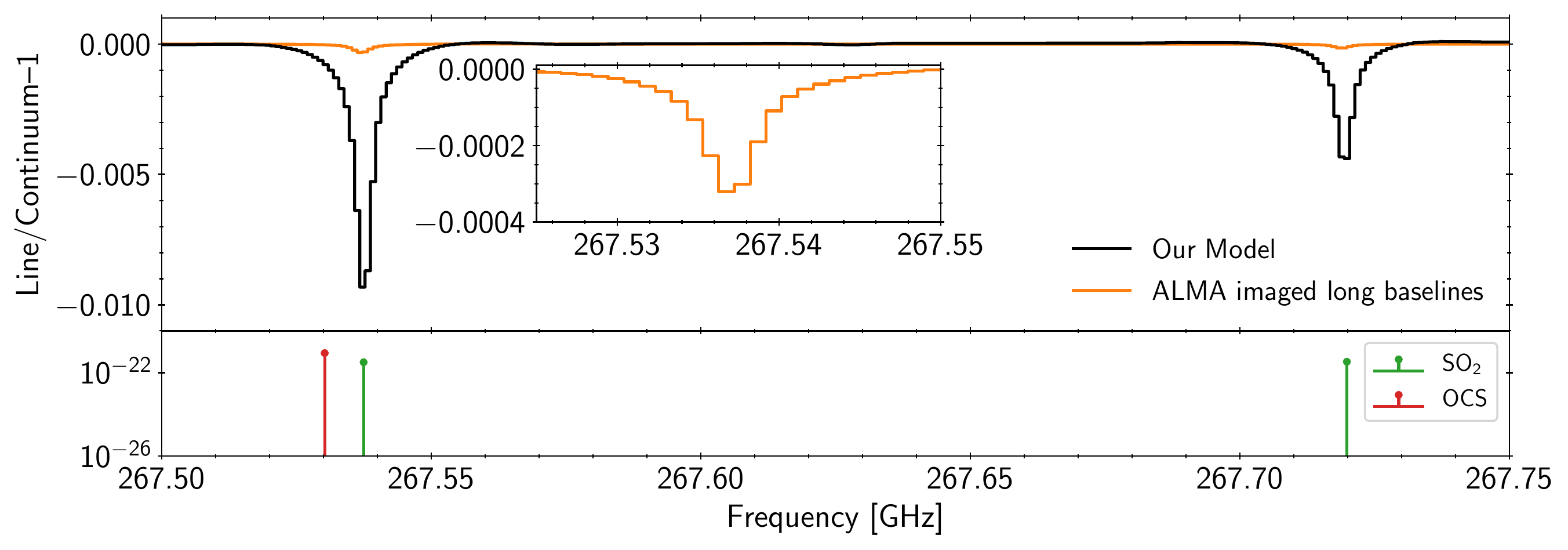}
    
    \caption{Modeled line dilution for the ALMA observations:  disk-averaged line/continuum ratios for our nominal Case D atmosphere model containing \ce{SO2} uniformly distributed over the Venus disk, at 0.00003~cm$^{-1}$ (1~MHz) resolution (black lines), for the detection frequency (left panel) and reference lines frequencies (right panel).  Absorption line strengths (lower panels) are given in units of cm$^{-1}$/(molecule~cm$^{-2}$). The orange lines show the same spectral model as imaged using the ALMA antenna configuration of the \citet{Greaves2020} observations. The inset shows the suppression of the 267.54 GHz reference line to an l:c of close to $-0.0003$.   The line intensity is significantly diluted, and is consistent with the non-detection of these \ce{SO2} absorption lines in the wideband data.}
    \label{fig:dilution}
\end{figure*}

While the non-detection of prominent \ce{SO2} spectral features in the ALMA wideband data could indicate a low abundance, as argued by \citet{Greaves2020}, the estimation of this abundance was done without correcting for line dilution as a result of the ALMA observing geometry \citep{Greaves2020}. Significant line dilution is likely, especially considering the global distribution of \ce{SO2} in the Venus atmosphere, and the exclusion of the short baseline ALMA measurements. \citet{Greaves2020} estimated line dilution (filtering losses) of 60--92\% depending on position on the disk. To determine the disk-averaged line dilution for the \ce{SO2} reference line search, we simulated observations of Venus using the ALMA configuration of \citet{Greaves2020} by imposing an appropriate resolution spectrum (0.00003~cm$^{-1}$, 1~MHz) of our Case~D atmospheric model over a limb-darkened disk model. The Fourier Transform of this model was re-sampled to match the ALMA configuration and re-imaged using the imaging routines of \citet{Greaves2020}, as provided in their Supplementary Software~3. As shown in Figure~\ref{fig:dilution}, line dilutions on the order of 95\% at the line core are observed for the full disk. We observe similar dilutions when the spectrum is only imposed on one hemisphere ($\sim$8~arcsecond extent at the time of observation). This line dilution suggests that the \ce{SO2} reference features produced by our best fit \ce{SO2} distribution (Case D) would be heavily suppressed by line dilution in the ALMA data, which would cause them to mimic smaller features below the ripple detection limit of $-0.0006$.

\section{Discussion}

The claim that \ce{PH3} has been detected in the Venus clouds is currently supported by observations of a single absorption line at a frequency that also coincides with absorption from \ce{SO2}, a known and relatively common Venus gas, and based on an emission weighting function that peaks at 56~km \citep{Greaves2020}. However, our radiative transfer analysis indicates that the line at 266.94~GHz does not measure absorption within the Venus clouds. Our explicit calculation of radiance Jacobians confirms the assessment that both 266.94~GHz \ce{PH3} and \ce{SO2} line core absorption would be produced well above the Venus cloud deck at altitudes exceeding 80~km. Arguments for a mesospheric origin for the 266.94~GHz line core, based on the observed narrow width of the line, are also provided in a recent commentary by \citet{Villanueva2020}. This mesospheric contribution is inconsistent with a vertical abundance profile that concentrates \ce{PH3} in the middle and upper clouds, as used by \citet{Greaves2020} and \citet{Bains2020} to interpret their discovery. Our spectral simulation using this photochemical \ce{PH3} profile also shows that it is not consistent with the strength of the observed 266.94~GHz line. However, the presence of \ce{PH3} in the Venus clouds is not conclusively ruled out either, a point also made by \citet{Greaves2020MA}, because the \citet{Greaves2020} observations are not sensitive to absorption at cloud deck altitudes, and so can neither exclude, nor confirm, the presence of \ce{PH3} in the Venus clouds.  

Given that we have shown that the observed 266.94~GHz line predominantly originates high in the mesosphere, attributing it to \ce{PH3} is less chemically plausible than \ce{SO2}.  At these higher altitudes ($>$80~km) \ce{PH3} would be destroyed rapidly, while \ce{SO2} is photochemically regenerated \citep{sandor2010sulfur,belyaev2012vertical,Zhang2012}. Between 82~km and 96~km (70--300~Pa, where the line core absorption originates, Fig.~\ref{fig:jacobians}) \ce{PH3} has a sub-second lifetime, due to the destruction by Cl and H radicals and UV photolysis \citep{Bains2020}. To balance this rapid destruction rate and maintain a mesospheric concentration of 20~ppb, an extremely large flux of \ce{PH3} is required, potentially as large as $3.7\times10^{15}$~molecules~cm$^{-2}$~s$^{-1}$. For comparison, this production rate is about $\sim$100 times the flux of \ce{O2} produced by Earth's global photosynthetic biosphere \citep{Field1998}, the dominant metabolism on our planet. \citet{Greaves2020}, assuming the 266.94~GHz absorption was from \ce{PH3} in the clouds, calculated a significantly smaller production rate of $10^{7}$~molecules~cm$^{-2}$~s$^{-1}$, due to the lower destruction rate within the clouds.  However, the assumption of this in-cloud production rate results in a \ce{PH3} mixing ratio that effectively falls to zero at $>$80~km altitude \citep[][Fig. 5b]{Greaves2020}, which is inconsistent with our analysis that the observed line is sourced in the mesosphere. Although a recent reanalysis of the ALMA data by \citet{Greaves2020MA} has greatly reduced the significance of the 266.94~GHz line detection, their assignment of 1~ppb of \ce{PH3} in the mesosphere would still require a production rate significantly higher than the Earth's photosynthetic biosphere, and the larger 20~ppb \ce{PH3} value inferred from the JCMT data still stands. 

These challenges to mesospheric production rate are not relevant if the observed 266.94 GHz line is instead attributed to \ce{SO2}, which is known to increase in abundance with altitude in the mesosphere \citep{belyaev2012vertical,mills2018simulations}. A combination of infrared observations that probe the upper cloud and lower mesosphere, and UV occultation measurements that probe the upper mesosphere, has been used to map the vertical distribution of mesospheric \ce{SO2} \citep{belyaev2012vertical,Belyaev2017}. This distribution drops from the cloud tops to a minimum just below 80~km, but increases substantially from 80--100~km to typically several hundred ppb \citep{belyaev2012vertical,vandaele2017sulfur}. 

Assuming that the Venus atmosphere does not contain \ce{PH3}, we find that a realistic vertical profile for \ce{SO2} fits the JCMT 266.94~GHz detection.  Because the JCMT observations were single dish, any \ce{SO2} contribution to the 266.94~GHz line would not have been suppressed, as was the case for the ALMA data, and so should be sensitive to the true mesospheric \ce{SO2} abundance.  We used a mesospheric \ce{SO2} profile that is based on the profile observed in 2007--2008 by \citet{belyaev2012vertical}, which is likely a good fit to similar higher values seen in 2016--2018, a time span that includes the \cite{Greaves2020} JCMT observation.  This profile is also consistent with cloud top values of 200--350 ppb observed in the mid-infrared within a month of the \citet{Greaves2020} JCMT observations \citep{Encrenaz2019hdoso2}. The \citet{Encrenaz2019hdoso2} observations support the validity of our \ce{SO2} vertical profile, and suggest that the Venus mesosphere was unlikely to be experiencing a period of anomalously low  \ce{SO2} abundance at the time of the JCMT observations. 
Using this vertical abundance profile and a \ce{CO2}-broadened \ce{SO2} line profile, we can fit the width and shape of the 266.94~GHz line using \ce{SO2} alone, without needing an additional \ce{PH3} component.  The \ce{SO2} is also a better fit to the line centroid than the \ce{PH3} (cf. Fig. \ref{fig:spectra} A/B, D).  This excellent fit counters the argument of \citet{Greaves2020MA} that \ce{SO2} alone would be too narrow to fit the observed line. \citet{Greaves2020MA} also recently argued that the \ce{SO2} abundance required to fit the JCMT 266.94~GHz line (evenly-mixed 150~ppb for their fit, and 100~ppb for \citealt{Villanueva2020}) is unrealistically large, given previous mm-wave observations, which have returned lower values for mesospheric \ce{SO2} \citep{sandor2010sulfur,Encrenaz2015venus}.  However, mm-wave observations do not have as long, or as well sampled, a baseline as dedicated Venus spacecraft observations of the mesosphere \citep{belyaev2012vertical,Belyaev2017,vandaele2017sulfur}, and mesospheric \ce{SO2} abundance has been observed to vary by an order of magnitude on daily to yearly timescales, with values at 90--95 km altitude between 10 to 300 ppb. There is also evidence for longer-term secular changes in mesospheric and cloud-top \ce{SO2} abundances, with maxima in 2007--2008 and 2016--2018, and a minimum in 2012--2014 \citep{Belyaev2017,Encrenaz2019hdoso2}. We note that the model that we used to fit the JCMT 266.94 GHz line assuming a higher abundance of \ce{SO2} also produced an accurate fit to the lower abundance observation of \citet{Encrenaz2015venus} (see our Fig. \ref{fig:encr}), which was observed near an \ce{SO2} minimum.

We also find that strong ALMA line dilution allows the vertical abundance profile of \ce{SO2} that fits the JCMT 266.94~GHz observations to still be consistent with the non-detection of the \ce{SO2} ALMA reference lines---which are likely poor indicators of the impact of \ce{SO2} on the JCMT observations. Spectral simulations using our Case~D \ce{SO2} vertical distribution predict \ce{SO2} lines at 267.54 and 267.72~GHz with l:c ratios that are close to a factor of 10 larger than the nominal ALMA non-detection limit of $-0.0006$ given by \cite{Greaves2020}. This apparent contradiction can be reconciled by the lack of line-dilution in the JCMT observation of the 266.94~GHz line, as the single-dish integrates flux over all scales, while the telescope configuration and the removal of measurements from the $\leq33$~m ALMA baselines would have likely resulted in at least 90--95\% line dilution (factor of 10--20 suppression) for spatially-uniform \ce{SO2} gas. Therefore, taking the sensitivity of the two telescopes into account, our JCMT fit does not need to be adjusted, but our modeled \ce{SO2} l:c ratios should be divided by at least $\sim$20, if the \ce{SO2} is uniform across the disk, to approximate the ALMA detection for that set of baseline configurations. In doing so, our predicted \ce{SO2} reference line values fall below the ``10 ppb" ($-0.0006$) detection threshold (see Figure~\ref{fig:dilution} inset). Consequently, the \ce{SO2}-only model with up to several hundred ppb of \ce{SO2} in the mesosphere can fit the JCMT data, and still be consistent with the non-detection of \ce{SO2} in the ALMA wide-band data. Moreover, this strong line dilution, with the corresponding loss of sensitivity to even high levels of \ce{SO2}, suggests that the ALMA wide-band \ce{SO2} reference observations were likely poor indicators that \ce{SO2} was low enough to be ruled out as a significant source of the JCMT 266.94 GHz line---thereby significantly weakening the argument that this line was instead due primarily to \ce{PH3}.

In addition to explaining the JCMT single-dish detection of the 266.94~GHz line, and the suppression of the \ce{SO2} reference lines in the ALMA data, our \ce{SO2}-only hypothesis would also predict that the 266.94 GHz ALMA line would be, like the \ce{SO2} reference lines, strongly suppressed by line dilution and potentially non-detectable.  While this was not the case in the original \citet{Greaves2020} paper, this is now consistent with recent significant challenges to the detection confidence of the 266.94~GHz ALMA line.  These include  reanalyses of the \citet{Greaves2020} narrowband ALMA discovery data by both \citet{Snellen2020alma} and \citet{Villanueva2020} who concluded that the feature attributed to \ce{PH3} could not be detected with statistical significance. Our own further analysis of the \citet{Greaves2020} ALMA data, including testing the robustness of the detection at 266.94~GHz, comes to a similar conclusion, and is presented in \citet{Akins2020}. Additionally, a recent reanalysis of high-resolution, S/N$\sim$1000 Venus observations taken in 2015 was used to search for a \ce{PH3} transition near 10.47~\um{}, but it was not detected, setting a stringent upper limit of 5~ppb above the Venus clouds \citep{Encrenaz2020}. Finally, the recent \citet{Greaves2020MA} communication analyzing a reprocessing of the ALMA data suggests that the 266.94 GHz feature in the narrow-band whole-planet ALMA data is now significantly reduced in detection significance from the original discovery paper (4.8-$\sigma$ vs 13.3-$\sigma$), with an l:c of $-2\times10^{-5}$, consistent with 1~ppb of \ce{PH3}.  However, this much-reduced 266.94~GHz feature would also be consistent with line-diluted \ce{SO2}, which in our model would have l:c of $-1$ to $-2\times10^{-5}$ at this frequency, for line dilution in the range 95--97\%---which is likely well within the range of potential line dilution \citep{Akins2020}.

Although the \ce{SO2} hypothesis self-consistently explains our current understanding of the detection and non-detections in the JCMT and ALMA data, additional analyses and observations will be needed to more definitively discriminate between \ce{PH3} and \ce{SO2} as the source of the 266.94~GHz JCMT line.   Re-observing Venus at 266.94~GHz will likely still be needed to independently confirm the discovery observation, and detection of an additional \ce{PH3} absorption feature would provide a much stronger case for its presence in the Venus atmosphere. Future observations to confirm the \ce{PH3} $J=1\leftarrow0$ line detection should incorporate single dish measurements, which would not suffer from line dilution, or observations including the Atacama Compact Array (which includes shorter baseline measurements than the primary ALMA array). Because the \ce{SO2} abundance is critical to the \ce{PH3} identification for the ALMA data, we recommend that future attempts to confirm the ALMA \ce{PH3} observations should also obtain near-simultaneous \ce{SO2} measurements. The narrowband correlator configuration can be tuned to 266.94~GHz and to the frequencies of two nearby, stronger \ce{SO2} lines (near 267.54 and 267.72~GHz). To mitigate the spectral ripple features that compromised measurement of the line intensities \citep{Greaves2020}, these observations should occur when the apparent angular diameter of Venus is smaller and therefore less resolved by the ALMA antennas. 

Ultimately, the claimed detection of \ce{PH3} in the atmosphere of Venus has underscored the necessity of identifying and assessing the context of the environment within which we find potential biosignatures. The identification of the 266.94~GHz line as due to \ce{PH3}, and its plausibility as a potential biosignature, is inextricably intertwined with the physical and chemical environment of the Venus cloud and above-cloud atmosphere. This initial, controversial detection has highlighted just how much we still need to understand about our sister planet, and how important that knowledge is in interpreting this discovery. If the 266.94~GHz line is confirmed, and conclusively attributed to \ce{PH3}, its presence in the mesosphere would require additional observations to understand potential sources and sinks, and the attendant (and as yet unknown) phosphorous chemistry that enables its persistence at these high altitudes. Moreover, if \ce{PH3} is being generated abiotically, especially at these high altitudes, this would have negative implications for the robustness of \ce{PH3} and other reduced gases to serve as biosignatures in oxidizing terrestrial atmospheres. Regardless of the outcome, additional targeted observations will reveal processes on a terrestrial planet that informs our understanding of our own world, and potentially a large number of exoplanets that may share a similar evolutionary path and current environment.

\section{Conclusions}

We simulated millimeter-wavelength Venus spectra to explore the vertical distribution and detectability of \ce{PH3} and \ce{SO2} in the Venus atmosphere.  We find that the observations of the 266.94~GHz absorption line are insensitive to the abundance of \ce{PH3} and \ce{SO2} within the cloud deck.  Instead, the observed absorption at this wavelength originates from the mesosphere at altitudes above 80~km.  At these altitudes, \ce{PH3} would be rapidly destroyed, such that $20\pm10$~ppb of \ce{PH3} would require a flux of \ce{PH3} to the Venus mesosphere that is $\sim$100 times higher than the global production rate of photosynthetically-generated \ce{O2} on Earth.  Because \ce{PH3} and \ce{SO2} both absorb within the width of the line detected at 266.94~GHz, we emphasize that the identification of this absorption line as due to \ce{PH3} in both the ALMA and JCMT data relies heavily on the apparent low abundance of \ce{SO2} inferred from the non-detection of an \ce{SO2} reference line at 267.54~GHz in the ALMA data. However, we show that \ce{SO2} absorption is likely heavily suppressed in the ALMA data. Using \ce{SO2} vertical profiles within the range of previous observations (from 30~ppb at 78~km to 400$\pm$150~ppb at 100~km)---including \ce{SO2} observations taken within a month of the JCMT data---our model can fit the depth and width of the 266.94~GHz feature without \ce{PH3}.  We also show that ALMA line dilution suppresses the values for nominal Venus mesospheric \ce{SO2} to below the corresponding detectability limit set by \cite{Greaves2020}. Given the mesospheric altitude range, short chemical lifetime of \ce{PH3}, and consistency with existing mesospheric \ce{SO2} abundances observed within a month of the JCMT observations, we argue that \ce{SO2} provides a more self-consistent explanation for the 266.94~GHz feature than \ce{PH3}. Single dish observations optimized for Venus and used to assess the \ce{PH3} detection and \ce{SO2} abundance in the Venus upper mesosphere should be prioritized to discriminate between \ce{PH3} or \ce{SO2} as the source of the 266.94~GHz line.

\section{Acknowledgements}

This work was performed by the  Virtual Planetary Laboratory Team, a member of the NASA Nexus for Exoplanet System Science, and funded via NASA Astrobiology Program Grant No. 80NSSC18K0829. Part of this work was conducted at the Jet Propulsion Laboratory, California Institute of Technology, under contract with NASA. Government sponsorship acknowledged. This work made use of the advanced computational, storage, and networking infrastructure provided by the Hyak supercomputer system at the University of Washington.  We thank Jacob Lustig-Yaeger, Kevin Zahnle, and the anonymous reviewers for helpful comments. 

\software{LBLABC \citep{Meadows1996}, SMART \citep{Meadows1996}, CARTA \citep{Comrie2020carta}, CASA \citep{McMullin2007casa}, Matplotlib \citep{Hunter2007}, Numpy \citep{Walt2011}, GNU Parallel \citep{Tange2011}, WebPlotDigitizer \citep{Webplotdigitizer}.}

\bibliographystyle{aasjournal} \bibliography{VenusPH3}

\end{document}